\documentstyle[aps,prl,twocolumn,epsf]{revtex}

\begin{document}
\draft
\title{Realization of Bose-Einstein condensation of dilute magnons
in TlCuCl$_3$}

\author{T. Nikuni, M. Oshikawa, A. Oosawa and H. Tanaka}

\address{
Department of Physics, Tokyo Institute of Technology,
Oh-okayama, Meguro, Tokyo 152-8551, Japan}

\date{\today}

\maketitle

\begin{abstract}
The recent observation
[Oosawa et al. J. Phys. : Condens. Matter {\bf 11}, 265 (1999)]
of the field-induced N\'{e}el ordering in a spin-gap magnetic compound
TlCuCl$_3$ is interpreted as a Bose-Einstein Condensation of magnons.
A mean-field calculation based on this picture
is shown to describe well the temperature dependence of the magnetization.
The present system opens a new area for studying Bose-Einstein
condensation of thermodynamically large number of particles
in a grand-canonical ensemble.
\end{abstract}


The Bose-Einstein Condensation (BEC) is one of the most exotic
phenomena predicted by quantum mechanics~\cite{bec}.
Although the superfluid transition of Helium 4 may be regarded
as a BEC, it is influenced strongly by the interaction and
is much different from condensation of ideal or dilute Bose gas.
Recently, there has been a renewed interest in BEC, because
the realization of BEC by ultracooling of dilute atoms has
become possible~\cite{anderson}.
While the BEC of ultracooled atoms is of great interest,
there are various experimental limitations.
On the other hand, it has been known for a long time that
a quantum spin system can be mapped to an interacting Bose gas, and
that the off-diagonal long-range order which characterizes BEC
corresponds to a long-range magnetic order in the spin
system~\cite{bosemap}.
It is then possible to tune the density of bosons (magnons) by a
magnetic field to observe BEC of dilute bosons.
However, such an attempt has been apparently lacking.

In this letter, we argue that BEC of dilute bosons
in a thermodynamic number $\sim 10^{20}$
is realized in a recent high-field experiment on TlCuCl$_3$~\cite{Oosawa},
which is composed of chemical double chain of
Cu$_2$Cl$_6$ \cite{Oosawa,Takatsu}.
The compound has an excitation gap $\Delta/k_{\rm B} \approx 7.5$K above the
singlet ground state, in the absence of the magnetic
field \cite{Oosawa,Shiramura,Tanaka}. 
The origin of the gap may be attributed to the antiferromagnetic
dimer coupling in the double chain.
When the external field $H_{g} =  \Delta / (g \mu_{\rm B})$ to the gap
is applied, the gap collapses.
At finite temperature, the ``collapse'' of the gap at
$H_{g}$ does not give a singularity
because thermal excitations exist even if $H < H_{g}$.
However, there seems to be a phase transition due to the 
interchain interactions at higher field $H=H_c > H_g$, 
which depends on the temperature. 
In Ref.~\cite{Oosawa} the phase transition was identified as
a long-range magnetic ordering, and
was compared with
a mean-field theory (MFT)~\cite{TY,TY2} based on a dimer model.
While the dimer MFT does predict the field-induced
ordering, the experimental features were not well reproduced.
In particular, it predicts almost flat dependence of the
critical temperature $T_c$ on the magnetic field, while
in the experiment $T_c$ depends on the magnetic field by
a power law $T_c^{\phi} \sim H - H_g$ (see Fig.~1).
Moreover,
it predicts almost constant magnetization for $T<T_c$ and
concave magnetization for $T> T_c$, as a function of temperature $T$.
However, in the experiment, magnetization was found to increase
as decreasing $T$ below $T_c$ and it is a convex function of $T$
for $T> T_c$ (see Fig.~2).

We will show that the transition is rather well described as
BEC of magnons.
While the details of exchange interactions in TlCuCl$_3$
are not known yet, excitations above the singlet ground state
generally can be treated as a collection of bosonic particles
-- magnons~\cite{chub}.
If the exchange interaction is isotropic, which seems to be the
case in TlCuCl$_3$, the number of magnons are conserved
in a short timescale (but not conserved in a longer timescale.)
We assume that magnons carry spin $1$, as generally expected.

Under a magnetic field $H \sim H_{g}$,
the magnons with $S^z=1$ can be created by small energy.
Thus, at low temperatures $T \ll \Delta$ and $H \sim H_g$,
we can consider only those magnons.
The chemical potential of the magnons are given by
$\mu = g\mu_{\rm B} (H - H_g)$.
The total number of magnons $N$ is associated with the total magnetization $M$
through $M=g\mu_{\rm B}	N$.
If the magnons are free bosons, the number of magnons would be
infinite for $H > H_g$.
However, in the spin system, magnons cannot occupy the same sites and thus
there is a hard-core type interaction between them.
The interaction keeps the number of magnons to be finite.

The transverse components of the exchange interactions give rise to 
hopping of the magnons, while the longitudinal component give rise to
the interaction.
Although the exchange interaction and thus hopping
might be complicated, generically the dispersion relation of a magnon
is quadratic near the bottom.
Thus the low-energy effective Hamiltonian for the ($S^z=1$) magnons
are given by
\begin{eqnarray}
H &\sim& \sum_k 
\left[ \left( \sum_{\alpha=x,y,z} \frac{\hbar^2k_{\alpha}^2}{2 m_{\alpha}} \right) - \mu 
\right]
a^{\dagger}_k a_k \cr
&& + \frac{1}{2} \sum_{k,k',q} v({\bf q})
	a^{\dagger}_{k+q} a^{\dagger}_{k'-q} a_{k} a_{k'} + \ldots
\end{eqnarray}
Here the momentum ${\bf k}$ is measured from the minimum of the
magnon dispersion.
For simplicity, we do not consider the case where the magnon dispersion has
more than one minimum~\cite{NS}.
The effective masses $m_{\alpha}$ is related to the curvature
of the dispersion relation in the direction of $\alpha$.
By a rescaling of momentum, we may consider isotropic effective
Hamiltonian instead.
This is nothing but the non-relativistic bosons
with a short-range interaction.

Moreover, in the low-density and low-temperature limit,
only the two-particle interaction is important and it can be
replaced by delta-function interaction $v(q) \sim v_0$.
Thus the effective Hamiltonian is given by
\begin{equation}
H = \sum_k 
\left( \frac{\hbar^2k^2}{2 m} - \mu \right)
a^{\dagger}_k a_k
+ \frac{v_0}{2} \sum_{k,k',q}
	a^{\dagger}_{k+q} a^{\dagger}_{k'-q} a_{k} a_{k'}.
\label{eq:boseham}
\end{equation}
This effective Hamiltonian can be derived from some specific
models~\cite{Kol,GT}.
However, we emphasize that it is universal at the low-temperature
and low magnon density limit, and does not
depend on details of the exchange interaction.

Since the number of magnon is actually not conserved due to
the small effects neglected in the Hamiltonian, we have
a grand canonical ensemble of the bosons.
The ``chemical potential'' can be controlled precisely by 
tuning the magnetic field.
When the chemical potential becomes larger than a critical value,
the system undergoes a BEC.
Thus the spin-gap system in general would provide a great opportunity
to study BEC in a grand canonical ensemble, with thermodynamically
large number of particles.

The idea that BEC is induced by the magnetic field in a spin-gap
system has appeared several times.
There was a discussion of (quasi-) Bose condensation 
in a Haldane gap system under a magnetic field~\cite{Affleck:bosecon}.
However, there is no BEC at finite temperature in a one-dimensional system.
On the other hand, the experiments on Haldane gap systems are often
affected by the anisotropy and the staggered $g$-tensor,
which wipe out the BEC.
Giamarchi and Tsvelik~\cite{GT} have recently discussed the
three-dimensional ordering in coupled ladders in connection with BEC. 
However, as far as we know, there has been no experimental
observation of the magnon BEC induced by an applied field.

We first consider the normal (non-condensed) phase.
Within the Hartree-Fock (HF) approximation, the momentum distribution of the
magnons is given by~\cite{FW}
\begin{equation}
n_k\equiv \langle a^{\dagger}_ka_k^{\dagger} \rangle
=\frac{1}{e^{\beta(\varepsilon_k-\mu_{\rm eff})}-1},
\end{equation}
with $\varepsilon_k\equiv\hbar^2k^2/2m$ and $\mu_{\rm eff}\equiv\mu-2v_0n$.
The magnon density $n\equiv N/N_d$ 
($N_d$ is the total number of the dimer pairs)
has to be determined self-consistently by
\begin{equation}
n=\int\frac{d^3k}{(2\pi)^3}n_k=\frac{1}{\Lambda^3}g_{3/2}(z),
\label{eq:n}
\end{equation}
where $z\equiv e^{\beta \mu_{\rm eff}}$ is the fugacity,
$\Lambda\equiv(2\pi\hbar^2/mk_{\rm B}T)^{1/2}$ is the thermal de Broglie wavelength,
and $g_n(z)\equiv\sum_{l=1}^{\infty} z^l/l^n$ is the Bose-Einstein function.
BEC occurs when the effective chemical potential $\mu_{\rm eff}$ vanishes
so that $\mu=2v_0n$.
Setting $z=1$ in (4) gives the temperature dependence of the critical
value of the chemical potential
\begin{equation}
\mu_c=2v_0\left(\frac{mk_{\rm B}T}{2\pi\hbar^2}\right)^{3/2}\zeta(3/2).
\label{eq:muc}
\end{equation}
This implies that the temperature dependence of the critical magnetic field
at low temperatures is $H_c(T)-H_g\propto T^{3/2}$.
This power-law dependence is independent of the interaction parameter $v_0$.

When $\mu>\mu_c$, one has the macroscopic condensate order parameter
$\langle a_0 \rangle=\sqrt{N_c}e^{i\theta}\neq 0$, where $N_c$ is the total number
of the condensate magnons.
In terms of the original spin system, this means that
there is a (staggered) transverse
magnetization component $m_{\perp}=g\mu_{\rm B}\sqrt{n_c/2}$ with
$n_c\equiv N_c/N_d$.
Within the Hartree-Fock-Popov (HFP) approximation, 
the condensate density is determined by \cite{griffin}
\begin{equation}
\mu = v_0 n_c + 2 v_0\tilde n,
\end{equation}
where $\tilde n=n-n_c$ is the density of the non-condensed magnons,
which is given by
\begin{eqnarray}
\tilde n&=&
\int\frac{d^3k}{(2\pi)^3}\left[
\left(\frac{\varepsilon_k+v_0n_c}{2E_k}-\frac{1}{2}\right)+
\frac{\varepsilon_k+v_0n_c}{E_k}f_{\rm B}(E_k)\right] \cr
&=&\frac{1}{3\pi^2}\left(\frac{mv_0n_c}{\hbar^2}\right)^{3/2}+
\int\frac{d^3k}{(2\pi)^3}
\frac{\varepsilon_k+v_0n_c}{E_k}f_{\rm B}(E_k),
\end{eqnarray}
where we have used the HFP energy spectrum 
$E_k=\sqrt{\varepsilon_k^2+2\varepsilon_kv_0n_c}$ and the
Bose distribution $f_{\rm B}(E_k)=1/(e^{\beta E_k}-1)$.
The first term of (7) represents the depletion of the condensate due to interaction
between magnons, which reduces to the ground-state non-condensate density at $T\to 0$.
The second term is the contribution from thermally excited non-condensate magnons,
which vanishes at $T\to 0$.
Eq.(6) is to be solved self-consistently in conjunction with Eq.(7).
Then the total magnon density is given by
\begin{equation}
n=n_c+\tilde n=\frac{\mu}{v_0}-\tilde n.
\end{equation}
In particular, the magnon density at $T\to 0$ is given by
\begin{equation}
n\approx\frac{\mu}{v_0}+\frac{1}{3\pi^2}\left(\frac{m\mu}{\hbar^2}\right)^{3/2}.
\end{equation}

If we ignore the deviation of $z$ from $1$ for $T \neq T_c$,
we obtain a simple result (using $v_0n_c=0$ in (7) for $T<T_c$)~\cite{GT}:
\begin{eqnarray}
\frac{n(T)}{n(T_c)} &=& 
\left(\frac{T}{T_c}\right)^{3/2} \ \  ( T > T_c), \cr
\frac{n(T)}{n(T_c)} &=& 
2 - \left(\frac{T}{T_c}\right)^{3/2} \ \ ( T < T_c).
\label{eq:simple}
\end{eqnarray}
Thus it predicts the cusp-like minimum of the magnon density (magnetization)
at $T=T_c$.
In contrast, the dimer MFT~\cite{TY} predicts a constant magnetization
below $T_c$.

Figure~\ref{fig:mexp} shows the observed low-temperature
magnetizations of TlCuCl$_3$ at various external fields for
$H \parallel b$. We can see the cusp-like anomaly at the
transition temperature, as predicted by the present theory.
The similar temperature dependence of the magnetization
can be observed for $H \perp (1,0,{\bar 2})$ \cite{Oosawa}. 
Thus the main qualitative feature of the temperature
dependence of the magnetization, which cannot be understood
in the dimer MFT, is captured by the magnon BEC picture.
The increase of $n$ for decreasing $T$ below $T_c$
is due to condensation of the bosons;
the cusp shape of the magnetization curve observed
in the experiment can be regarded as an evidence of the magnon BEC.
We note that, in the range of the experiment, the magnon density
is of order of $10^{-3}$ and is consistent with the assumption of
diluteness.

However, the approximation~(\ref{eq:simple}) does not precisely
reproduce the experimental result.
In particular, it predicts independence of $n$ on the applied field
$\mu$ for $T> T_c$ while the dependence was observed experimentally.
Part of the discrepancy may be due to the approximation $z=1$.
Actually, even in the HF framework, the approximation
$z=1$ cannot be justified.
In Fig.~3, we plot the temperature dependence of the total density $n$
above and below the transition temperature $T_c$ obtained by 
solving the self-consistency equations (4) and (8) numerically.
The interaction parameter $v_0$ and the effective mass $m$ are 
estimated from the experimental data as $v_0/k_{\rm B}\approx 400$K 
and $mk_{\rm B}/\hbar^2\approx0.025$K$^{-1}$.
The self-consistent calculation does predict the
total density $n$ dependent on the applied field for $T>T_c$,
which is qualitatively consistent with the experiment.
In Fig.~\ref{fig:nc} we also plot the temperature dependence
of the staggered transverse magnetization component $m_{\perp}$.
Direct measurements of $m_{\perp}$ using neutron diffraction are
in fact intended.

We see a discontinuity in magnon density (magnetization) at the transition point.
This is because our HFP approximation is inappropriate
in the critical region, and leads to an unphysical jump in the
condensate density
$n_c$ (for detailed discussion, see \cite{SG}).
In the vicinity of the critical point, the HFP approximation
eventually breaks down; the critical behavior
then belongs to the so-called 3D XY universality class~\cite{3DXY}.
On this ground, in the vicinity of $T_c$,
the transverse magnetization $m_{\perp}$ is expected as
$m_{\perp} \propto (T_c -T)^{\beta}$,
where $\beta \sim 0.35$.

\begin{figure}[ht]
 \epsfxsize=80mm
  \centerline{\epsfbox{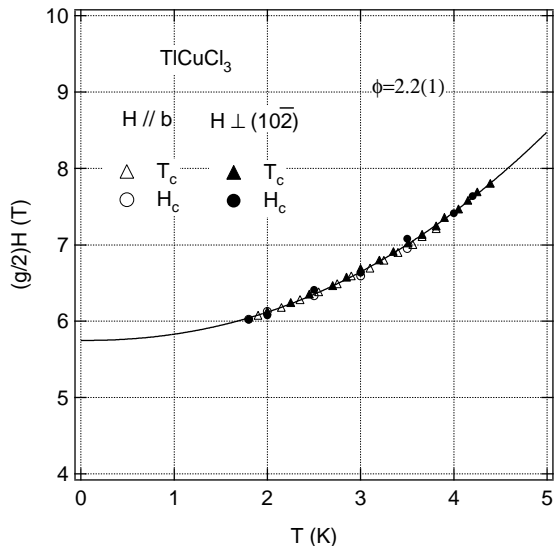}}
\caption{The phase diagram in TlCuCl$_3$.
The solid line denotes the fitting with Eq.(11) using 
$(g/2)H_c(0)=5.61$T and $\phi=2.2$.}
\label{fig:phase}
\end{figure}

\begin{figure}[ht]
 \epsfxsize=60mm
  \centerline{\epsfbox{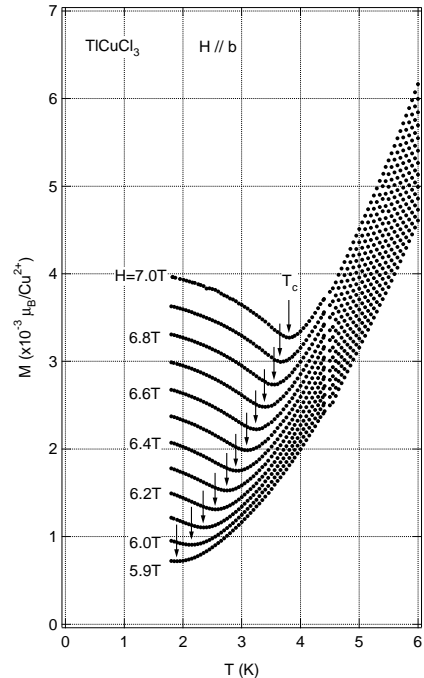}}
\caption{
The low-temperature magnetizations of TlCuCl$_3$ measured at various external
fields for $H \parallel b$.}
\label{fig:mexp}
\end{figure}

Figure~\ref{fig:phase} shows the experimentally determined
magnetic phase diagram of TlCuCl$_3$.
We fit the phase boundary $H_c$ as a function as a temperature $T_c$ with the following formula:
\begin{equation}
(g/2)[H_c(T)-H_c(0)]\propto T^{\phi}.
\end{equation}
The best fitting is obtained with $(g/2)H_c(0)=5.61$T and $\phi=2.2$~\cite{footnote}.
The obtained exponent $\phi=2.2$ disagrees somewhat with
the HF approximation~(\ref{eq:muc}) which gives $\phi=3/2$.
We note that, exactly $z=1$ holds at the transition point,
and thus $\phi=3/2$ is a definite conclusion within the
HF framework.
On the other hand, the dimer MFT predicts $H_c(T)$ to be
exponentially flat at low temperature~\cite{TY}.
The observed power-law dependence is qualitatively consistent with
the magnon BEC picture, compared to the dimer MFT.

As discussed above, our mean-field  analysis for a dilute Bose gas is not reliable
in the critical region, and thus the discrepancies with the
experiment may be attributed to the fluctuation effects.
More precise description of the experiment near the critical point
therefore requires the inclusion of the fluctuation effects.
Furthermore, in the experiment
there may be other effects that were ignored
in the effective Hamiltonian~(\ref{eq:boseham}), such as impurities.
These will be interesting problems to be studied in the future.

To conclude, we believe that the essential feature of the experimental
observation on TlCuCl$_3$, which cannot be understood in the traditional
dimer MFT, is captured by the magnon BEC picture.
The present system provides the first clear
experimental observation of field-induced magnon BEC,
with thermodynamically large number of particles.
It opens a new area of BEC research in grand canonical ensemble
with an easily tunable chemical potential (magnetic field).
Similar BEC of magnons would be observed in other magnetic materials
in the vicinity of the gapped phase, which may be
the singlet ground state due to large single-ion anisotropy \cite{tsune},
the completely polarized state \cite{bosemap,NS}, or the ``plateau'' phase
in the middle of the magnetization curve \cite{plateau}.
An essential requirement for observing BEC is that the system
has the rotationally invariance about the direction of
the applied magnetic field,
so that the number of magnons is (approximately) conserved.

We thank H. Shiba for useful comments.
T.N. was supported by JSPS and M.O. was supported in part
by Grant-in-Aid from Ministry of Education, Culture and Science
of Japan.


\begin{figure}[ht]
 \epsfxsize=80mm
  \centerline{\epsfbox{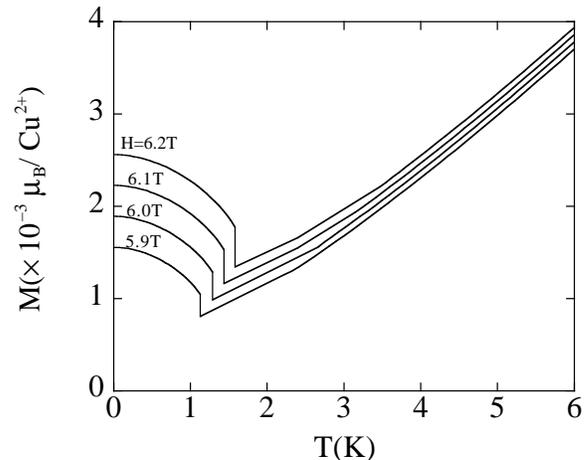}}
\caption{
The temperature dependence of the magnetization.
We have used $v_0/k_{\rm B}=400$K,  $mk_{\rm B}/\hbar^2=0.025$K$^{-1}$ and $g=2.06$.}
\label{fig:mtheo}
\end{figure}

\begin{figure}[ht]
\epsfxsize=80mm
\centerline{\epsfbox{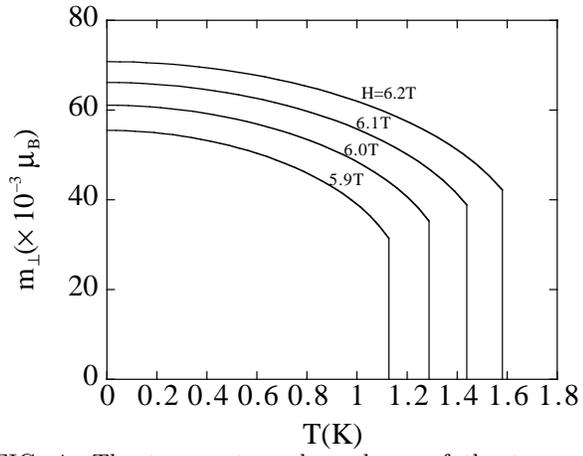}}
\caption{The temperature dependence of the transverse magnetization $m_{\perp}$.
We have used the same set of parameters as in Fig.3.}
\label{fig:nc}
\end{figure}


\begin{references}
\bibitem{footnote}
In Ref.~[4], temperatures with inflection points in the magnetization curve
are assign to the transition temperatures.
This analysis leads to a different exponent $\phi=1.7$.
\end{references}
\end{document}